\pdfoutput=1

\documentclass[sigchi-a,nonacm]{acmart}

\AtBeginDocument{%
  \providecommand\BibTeX{{%
    \normalfont B\kern-0.5em{\scshape i\kern-0.25em b}\kern-0.8em\TeX}}}

\setcopyright{acmcopyright}
\copyrightyear{2020}
\acmYear{2020}
\acmDOI{10.1145/1122445.1122456}

\acmConference[CSCW '20]{Ethics in Design CSCW Workshop}{October 2020}{Virtual Conference}
\acmBooktitle{Ethics in Design CSCW Workshop, October 2020, Virtual Conference}
\acmPrice{15.00}
\acmISBN{978-1-4503-XXXX-X/18/06}

\usepackage{xurl}
\usepackage{enumitem}
\setlist[itemize]{noitemsep, topsep=-10pt}

\usepackage[framemethod=tikz]{mdframed}
\definecolor{tealgreen}{rgb}{0.22, 0.75, 0.698}
\newmdenv[innerlinewidth=2pt,topline=false,rightline=false,bottomline=false,skipbelow={6pt},skipabove={8pt},linecolor=tealgreen,innerleftmargin=6pt,innerrightmargin=18pt]{mybox}

\begin{document}

\title{Explainability Case Studies}

\author{Ben Zevenbergen}
\affiliation{%
  \institution{Google}}
\email{benzevenbergen@google.com}

\author{Patrick Gage Kelley}
\affiliation{%
  \institution{Google}}
\email{patrickgage@acm.org}

\author{Allison Woodruff}
\affiliation{%
 \institution{Google}}
 \email{woodruff@acm.org}

\renewcommand{\shortauthors}{Zevenbergen, et al.}

\begin{abstract}

Explainability is one of the key ethical concepts in the design of AI systems. However, attempts to operationalize this concept thus far have tended to focus on approaches such as new software for model interpretability or guidelines with checklists. Rarely do existing tools and guidance incentivize the designers of AI systems to think critically and strategically about the role of explanations in their systems. We present a set of case studies of a hypothetical AI-enabled product, which serves as a pedagogical tool to empower product designers, developers, students, and educators to develop a holistic explainability strategy for their own products.
\end{abstract}



\maketitle

\section{Introduction}
Explainability has been highlighted as an important pillar of responsible AI practices~\cite{fjeld2020,jobin2019}.  
However, despite explainability being a focus of academic scholarship, technical development, industry guidelines,\footnote{  \url{https://www.blog.google/technology/ai/ai-principles/}, for example}
and regulatory attention, best practices are not yet established for creating explanations that benefit individuals, communities, and expert audiences. Constructing good explanations for AI systems is a complex and largely untested design issue that does not yet lend itself to checklists but rather calls for more open-ended exploration~\cite{woodruff2019}. We aim to support AI designers, developers, educators, and others in the challenging task of considering how and where to deploy clear, understandable explanations that improve outcomes for individuals and society. 

Explanations that come at the ``moment of decision'' are the current de-facto standard and best practice for AI explanations as implemented today. By contrast, in this work we take a broad view of explainability. This includes the exact, technical explanations that some AI systems provide at the moment of decision or inference---but also covers general introductions to an AI system, reasons given in moments of failure or error, descriptions of personalization and change over time, FAQ and help materials, design nudges and other less explicit information, results of audits and investigations, educational and public service campaigns, and any other support that helps users of a system understand the decisions made about them by AI.

In this short paper, we present a set of case studies we have designed to support semi-structured, nuanced discussions of how explanations can be designed and deployed. 
Case studies are a particularly fruitful methodology to facilitate discussion and thinking about a range of variables that may influence outcomes, and they are often used in teaching technology ethics~\cite{10.1145/3328778.3366825}. 
Instead of prescribing steps to take, our case study approach instills in participants a set of new perspectives that enables an approach to explainability beyond model interpretation and narrow in-the-moment notifications. The case studies highlight limitations of the status quo and encourage participants to explore a wider range of opportunities, challenges, and solutions than are commonly considered.

\section{Related Work}

Explainability as a concept has received much attention in academic, policy, and business literature~\cite{gilpin2018,mittelstadt2019,selbst2018intuitive,miller2019}. We present a snapshot of the literature due to space limitations. 

Our case studies resonate with several ideas from academic literature. For example, a tiered system of transparency (through explanations) is highlighted in papers by Kaminski, Pasquale, and Edwards and Veale~\cite{kaminski2019right,edwards2017slave,pasquale2010}. 
Our case studies also add to a growing body of resources for ethical AI.
Tactical support for applying ethical AI ideas in practice is available in resources such as the Markkula Center Ethics in Technology Practice Framework and Toolkit,\footnote{\begin{minipage}[t]{8cm}
\url{https://www.scu.edu/ethics-in-technology-practice/}
 \end{minipage}} the Omidyar Ethical OS Toolkit,\footnote{\url{https://ethicalos.org/}} and the Princeton Dialogues on AI and Ethics Case Studies.\footnote{\url{https://aiethics.princeton.edu/case-studies/}}

Explainability also plays a role in checklists and toolkits that are largely aimed at having developers and designers build more ethical AI systems including Google's People in AI Research Guidebook\footnote{\url{https://pair.withgoogle.com/}}
and Responsible AI practices,\footnote{
 \begin{minipage}[t]{7cm} 
   \url{https://ai.google/responsibilities/responsible-ai-practices/?category=interpretability} 
 \end{minipage}}
 IBM's AI Explainability 360,\footnote{\url{http://aix360.mybluemix.net/}}
and PwC's Explainable AI.\footnote{ \begin{minipage}[t]{9cm}\url{https://www.pwc.co.uk/audit-assurance/assets/explainable-ai.pdf} \end{minipage}}
Policymakers and regulators have also published guidelines and checklists focused on explainability. The EU's High-Level Expert Group on Artificial Intelligence presents a brief selection of questions\footnote{\begin{minipage}[t]{9cm}\url{https://ec.europa.eu/futurium/en/ai-alliance-consultation/guidelines} \end{minipage}}
and the British Information Commissioner's Office (ICO) dedicates several publications to this topic.\footnote{\begin{minipage}[t]{9cm}\url{https://ico.org.uk/media/2616433/explaining-ai-decisions-part-2.pdf} \end{minipage}}

\section{The Case Studies}

The explainability case studies presented in this paper are a pedagogical tool that are intended to be deliberated in a workshop setting. The cases intentionally include some questionable or problematic explainability practices. In each of five situations, workshop participants discuss how and why to improve the design of the AI systems and their explanations. 

This approach takes participants out of their day-to-day context and places them into a hypothetical situation where an existing (though incomplete) explainability strategy is re-thought from the ground up. Using this approach, participants engage with ideas that underlie explainability best practices so they may then apply them in their own work.

The case studies revolve around an imagined high tech new car---The Model-U---with audio and visual sensors for identification of passengers, a personalized entertainment system, a reliable self-driving service, including taking complete driving control  
on highways, and advanced assistance in parking lots. Each of these features and systems are explored in the case studies.

\section{Materials}

\begin{sidebar}
\fontsize{10pt}{12.5}\selectfont
Complete materials for using these \newline case studies with a group of participants are available at:
\newline\textbf{ \url{https://arxiv.org/abs/2009.00246}}
\end{sidebar}

We provide a readme which gives an overview of the case studies and guidance on how to run an in-person or virtual workshop. The slide deck contains an agenda for the activities, an introduction to the basic concepts, and an overview of the Model-U and its features. Each of the five situations has the written case, as well as discussion prompts to help participants work through the ethics and values questions the cases describe. Table 1 summarizes the main themes and a short description of each situation.

\section{Conclusion}

These materials are designed for audiences interested in the design and development of technology, including but not limited to practitioners, developers, user experience professionals, and undergraduate or graduate students. No specific background or expertise is required. We hope that people will find these discussions engaging and that they may apply the ideas in their own work designing and critiquing technology.

\clearpage

\begin{table*}
    \centering
    \begin{tabular}{ l p{12cm} p{9cm} }
    \bf \#      &    \bf Situation      &    \bf  Themes \\  \midrule
    
    1 & 
    A family enters their newly purchased Model-U and are confronted with its identification system for the first time. One person is identified correctly and another is not, and the associated explanations cause user frustration.
    & 
    \begin{minipage}[t]{\linewidth}
    \begin{itemize}
    \item Superfluous content
    \item Inappropriate timing
    \item Explaining errors and uncertainty
    \item Empowering user action 
    \end{itemize}
    \end{minipage}
    \\ \midrule

    2 & 
    A driver merges onto the highway but is not ready to relinquish control of the Model-U to the self-driving system. The system's response makes an already complicated situation more stressful.
    & 
    \begin{minipage}[t]{\linewidth}
    \begin{itemize}
    \item Awareness of the user's context
    \item Tone of explanation
    \item Timeliness of content
    \item Company response to user complaint about explanation
    \end{itemize}
    \end{minipage}
    \\ \midrule

    3 & 
    The car's entertainment system recommends music, but it is unclear how user feedback informs its choices. The driver realizes they don't know enough about how the system works.
    &   
    \begin{minipage}[t]{\linewidth}
    \begin{itemize}
    \item Providing meaningful feedback
    \item Scarce attention
    \item Incomplete mental models
    \item Seeking appropriate moments for feedback
    \end{itemize}
    \end{minipage}
    \\ \midrule

    4 &
    The Model-U is in a minor accident, and the driver receives a complex, formal explanation that is not meaningful to them. Investigation reveals the accident was caused by an adversarial attack, and the company's public response is not sufficiently reassuring.
    &
    \begin{minipage}[t]{\linewidth}
    \begin{itemize}
    \item Varying end-user needs 
    \item Investigation of a high-profile failure
    \item Public transparency
    \item Communicating remote possibility of errors  
    \end{itemize}
    \end{minipage}
    \\ \midrule

    5 & 
    The Model-U's traffic avoidance system leads to congestion in towns near highways. A local council organizes a stakeholder meeting which turns into a participatory design exercise. There is a gap between the information community members want so they can co-develop policy and what the company is willing or able to provide.
    &
    \begin{minipage}[t]{\linewidth}
    \begin{itemize}
    \item Community participation and feedback
    \item Providing information to the public
    \item Responsiveness to diverse information requirements
    \item Limitations on transparency    
    \end{itemize}
    \end{minipage}
    \\ \bottomrule
    
    \end{tabular}
    \caption{A summary of the situations and themes in the five case studies.}
    \label{tab:my_label}
\end{table*}

\clearpage


\bibliographystyle{ACM-Reference-Format}
\bibliography{references}


\end{document}